\documentclass[doublecol]{epl2}
\usepackage{amsmath}

\usepackage{amssymb}

\usepackage{amsbsy}

\usepackage{graphicx}

\title{KAM tori in 1D random discrete nonlinear Schr\"odinger model?}

\author{Magnus Johansson\inst{1,4} \and Georgios Kopidakis\inst{2,4}
\and Serge Aubry\inst{3,4}}
\shortauthor{M. Johansson \etal}

\institute{
  \inst{1} Department of Physics, Chemistry and Biology (IFM), Link\"{o}ping 
University, SE-581 83 Link\"{o}ping, Sweden\\
  \inst{2} Department of Materials Science and Technology, University of 
Crete, GR-71003 Heraklion, Greece\\
  \inst{3} Laboratoire L\'eon Brillouin, 
CEA Saclay, 91191 Gif-sur-Yvette, France\\
  \inst{4} Max Planck Institute for the Physics of Complex Systems, 
N\"{o}thnitzer Str. 38, D-01187 Dresden, Germany
}

\abstract{
We suggest that KAM theory could be extended for certain 
infinite-dimensional systems with purely discrete linear spectrum. 
We provide empirical arguments for the 
existence of square summable infinite-dimensional invariant tori 
in the random discrete nonlinear Schr\"odinger equation, appearing with a
finite probability for a given initial condition with sufficiently small
norm. Numerical support for the existence of a fat
Cantor set of initial conditions generating almost-periodic oscillations is 
obtained by analyzing (i) sets of recurrent trajectories over successively 
larger time scales, and (ii) 
finite-time Lyapunov 
exponents. The norm region where such KAM-like tori may exist
shrinks to zero when the disorder strength goes to zero and the localization
length diverges. 
}    

\begin{document}

\maketitle

 KAM theory  \cite{KAMref} predicts  that, under generic conditions, 
Hamiltonian systems with a finite number of degrees of freedom, $N$, 
close enough to an integrable limit exhibit {\em quasiperiodic} trajectories
which are dense on invariant $N$-dimensional tori
in a  $2N$-dimensional phase space. 
The $N$ fundamental frequencies of those trajectories 
depend on the torus. When some integer combination of 
the frequencies vanishes, resonance is 
obtained, and tori which are resonant or almost resonant generally  break up 
into chaotic trajectories
when the Hamiltonian is perturbed from the integrable limit. 
Since there are infinitely many possible resonances,
infinitely many gaps appear densely in phase space. 
As the volume of the gaps drops exponentially with the order of 
the corresponding resonance, the persisting tori form a fat
Cantor set ({\it i.e.}, of nonvanishing Lebesgue measure),
which goes to full measure at the integrable limit. 
There is no general extension of the KAM theory for infinite 
systems ($N=\infty$), except for some special models \cite{FSW86,Po90}. 
However, it is generally believed that most KAM tori  disappear when the 
dimension
of the dynamical  system is infinite (see, {\it e.g.}, \cite{Fro}). 

Simple empirical arguments confirm that KAM tori,  
which are spatially localized
(square summable, $\mathit{l_2}$) cannot survive 
when the spectrum of the linearized system is \textit{absolutely continuous} 
in some frequency interval.
This situation occurs, {\it e.g.}, for spatially periodic arrays of coupled 
anharmonic oscillators.
In such models, any hypothetical quasiperiodic solution with at least two 
incommensurate frequencies would generate harmonics densely on 
the real axis, overlapping with the interval of 
the absolutely continuous part.  
Thus, these harmonics would radiate energy towards infinity
so that the localized energy could not be conserved \cite{AS09}. 
(By contrast, simple {\em periodic} solutions may remain localized,  
forming intrinsic localized modes 
[``discrete breathers''] which may be dynamical attractors
for some initial conditions \cite{DBref}.)
However, these arguments do {\em not} hold when the linear spectrum is 
{\em purely discrete}, and it is known, {\it e.g.}, that spatially localized 
periodic solutions 
with frequencies {\em inside} the linear spectrum exist generically in 
systems with linear Anderson localization \cite{AF91,IDB}.

We consider here the one-dimensional random discrete nonlinear 
Schr\"odinger (DNLS) system 
(see, {\it e.g.}, \cite{Shep93, Mol98, PS07, KKFA08, SKKF09, FKS09}),
\begin{equation}
 i \dot{\psi}_n = (\epsilon_n+\chi|\psi_n|^2) \psi_n-C(\psi_{n+1}+\psi_{n-1}),
 \label{rdnls}
 \end{equation}
with integrable limits at $\chi=0$ (linear limit) and $C=0$ 
(anticontinuous limit).
The random onsite energies $\epsilon_n$ are uniformly distributed in 
the interval 
$[-W/2:W/2]$. (See \cite{Lahini} for a direct experimental realization 
of (\ref{rdnls}) with optical waveguide arrays.) Then  the linear spectrum 
($\chi=0$) is discrete
and the corresponding eigenstates are $\mathit{l_2}$
(exponentially localized).  
Bourgain  and Wang \cite{BW08} have proven that close enough to the 
linear limit, eq.~(\ref{rdnls}) exhibits {\em quasiperiodic} solutions 
corresponding to  
{\em finite-dimensional} tori 
in phase space. 
Since 
in finite systems, invariant tori with dimension $P<N$ 
generally have
zero measure,  we expect similarly the solutions found in \cite{BW08}
to have zero probability to occur in an infinite system. 
In this work, we provide empirical arguments, supported by numerical 
calculations, suggesting  that eq.~(\ref{rdnls}) may also sustain 
{\em infinite dimensional} invariant tori of 
{\em almost periodic} 
\footnote{An almost periodic function $f(t)$ 
can be written as an absolutely
convergent series, 
$ f(t) = \sum_n f_n e^{i\omega_n t}$, 
where the set of frequencies $\omega_n$ is {\em countable} and 
$\sum_n |f_n| < +\infty$. \cite{Bohr}
}
trajectories, which are $\mathit{l_2}$ 
and could be found {\em with nonvanishing probability} for an initial 
condition at small enough norm.
Fr\"ohlich, Spencer and Wayne proved the existence 
of such tori
\cite{FSW86}, 
but only in a special family of modified random DNLS-type systems not 
including 
eq.~(\ref{rdnls}).

Our empirical arguments can be summarized as follows:
We consider an arbitrary $\mathit{l_2}$ 
small-amplitude initial condition, corresponding to a distribution of 
excited linear modes 
coupled by weak nonlinearities. 
Estimating the probability  that these couplings induce resonances 
between the linear modes, we find
{\em this probability to vanish in the limit of small norm}. 
Since our argument explicitly uses the fact that the linear spectrum
is discrete with exponentially localized eigenstates, it does not hold 
for a system with an absolutely continuous part
in its spectrum. 
Moreover, we predict that the existence domain of KAM-like tori shrinks to 
zero 
in the limit of weak disorder, 
or equivalently long localization length.

For the linearized eq.~(\ref{rdnls}),  the
discrete (countable) set of  eigenvalues $\omega_p$
is associated with a basis of real 
$\mathit{l_2}$ eigenstates $\{\phi_n^{(p)}\}$: 
$ \omega_p \phi_n^{(p)} =  
\epsilon_n \phi_n^{(p)}-C(\phi_{n+1}^{(p)}+\phi_{n-1}^{(p)})$.
Expanding $\{\psi_n\}$ in this basis, 
$\psi_n(t) = \sum_p \mu_p(t)  \phi_n^{(p)}$, yields
$ |\psi_n(t)|^2 = \sum_{p,p^{\prime}} 
\mu_p^{\star}(t) \mu_{p^{\prime}}(t) \phi_n^{(p)} \phi_n^{(p^{\prime})}$,  and the 
norm square
$$ {\mathcal N} \equiv  \sum_n  |\psi_n|^2 = \sum_{p} |\mu_p|^2 , $$
which is the second conserved quantity of eq.~(\ref{rdnls}) (in addition 
to the Hamiltonian).
Then, we obtain for the new complex coordinates $\mu_p$ 
(cf, {\it e.g.}, \cite{Shep93,SKKF09,FKS09}),
\begin{equation}
 i \dot{\mu}_p = 
 \omega_p \mu_p +  \sum_{p^{\prime}}  C_{p,p^{\prime}} (t)\mu_{p^{\prime}}(t), 
\label{eq1}
\end{equation}
where
$C_{p,p^{\prime}} (t)
\equiv  \chi \sum_{q,q^{\prime}} \mu_q^{\star}(t) \mu_{q^{\prime}}(t) 
V_{p,p',q,q'}$ are defined via the overlap sums
$V_{p,p',q,q'} \equiv 
\sum_n  \phi_n^{(p)} \phi_n^{(p^{\prime})}\phi_n^{(q)} \phi_n^{(q^{\prime})}
$. 
The coefficients $C_{p,p^{\prime}} (t)$  are real and depend on time through 
the coordinates themselves.

  In the limit of small amplitude $|\mu_p|$, cubic terms in 
eq.~(\ref{eq1}) are higher order and may be neglected 
during some time. 
The linear behavior  is $\mu_p(t) \approx  \mu_p(0)e^{-i \omega_p t}$. 
Then during that interval of time, coefficients $C_{p,p^{\prime}}(t)$  are 
almost periodic functions of time
({\it i.e.}, with a countable set of periods), 
$C_{p,p^{\prime}} (t) \approx  \chi \sum_{q,q^{\prime}} 
V_{p,p',q,q'}
\mu_q^{\star}(0) \mu_{q^{\prime}}(0)  e^{i (\omega_q-\omega_{q^{\prime}}) t}
$.
From  eq.~(\ref{eq1}), $ \frac{d}{dt} |\mu_p|^2 
 = 2 \sum_{p^{\prime} \neq p}  C_{p,p^{\prime}} (t) \mathrm{Im} (\mu_p^{\star}
 \mu_{p^{\prime}})$, 
defining the norm current $J_{p^{\prime} \rightarrow p}$ 
between two different Anderson modes $p \neq p^{\prime}$  
as
\begin{eqnarray}
 J_{p^{\prime} \rightarrow p}= 2
 C_{p,p^{\prime}} (t) \mathrm{Im} (\mu_p^{\star} \mu_{p^{\prime}})
\nonumber
\\
\approx 2 \chi \sum_{q,q^{\prime}} 
\left[
V_{p,p',q,q'}
|\mu_q(0)| \cdot |\mu_{q^{\prime}}(0)|
\ e^{i \left((\omega_q-\omega_{q^{\prime}})t
-(\alpha_q-\alpha_{q^{\prime}})\right)} \right] 
\nonumber
\\
\times |\mu_p(0)| \cdot |\mu_{p^{\prime}}(0)|
\sin( (\omega_p-\omega_{p^{\prime}})t-(\alpha_p-\alpha_{p^{\prime}})) , 
\end{eqnarray}
where  $\alpha_p$ is the initial phase of 
$\mu_p(0)= |\mu_p(0) | e^{-i \alpha_p}$. \
This current is almost periodic in time, and its 
oscillations should be small in order that  $|\mu_p(t)|^2$ remains 
approximately constant, so that 
 $\mu_p(t) \approx  \mu_p(0)e^{-i \omega_p t}$ remains valid for all times.
Since time integration of the current yields denominators 
$\omega_q-\omega_{q^{\prime}} \pm (\omega_p-\omega_{p^{\prime}})$  which may be 
small, this condition requires that there should be no strong resonances 
between any pairs of sites $p \neq p^{\prime}$. We consider resonances 
to be weak 
enough 
when  the numerator is smaller than the denominator for all terms, 
{\it i.e.},
\begin{equation}
 | \omega_q-\omega_{q^{\prime}} \pm (\omega_p-\omega_{p^{\prime}})| \gtrsim 
\kappa \left| \chi 
V_{p,p',q,q'} 
\right| 
\cdot |\mu_q(0)|\cdot |\mu_{q^{\prime}}(0)|
\label{nonrescr}
\end{equation}
for 
all $q$ and $q^{\prime}$, where $\kappa$ is of order $1$ (Chirikov criterion).
Note also that $\pm$ can be dropped, since the same condition is
obtained
if $q$ and $q^{\prime}$ are exchanged.

We assume now in order to fix the ideas that $\omega_p$ are random numbers  
distributed  in some interval 
with a smooth probability law with maximum density
$P_0$. 
Each resonance $p,p^{\prime},q,q^{\prime}$  has thus a probability 
with the 
upper bound 
$2 P_0 \kappa \left| \chi V_{p,p',q,q'}  \right| \cdot
|\mu_q(0)|\cdot |\mu_{q^{\prime}}(0)|$ to occur. 
The probability $P_R$ that  there is {\em at least one resonance} 
in the system for this initial condition, is bounded by the sum of these 
probability  bounds divided by $2$ 
(resonance $p,p^{\prime},q,q^{\prime}$  is 
equivalent to $p^{\prime},p,q^{\prime},q$):  
$$P_R \leq 
P_0 \kappa  | \chi | \sum_{q,q^{\prime}}  
 |\mu_q(0)| A_{q,q^{\prime}} |\mu_{q^{\prime}}(0)| ,$$
where
$A_{q,q^{\prime}} \equiv  \sum_{p\neq p^{\prime}}  | V_{p,p',q,q'} | $.
Then we obtain
$$P_R  \leq  P_0 \kappa |\chi| \cdot ||\mathbf{A}|| \cdot ||\{\mu_q(0)\}||^2 ,$$
with $\mathbf{A}= \{A_{q,q^{\prime}}\}$ and the norm
$$||\mathbf{A}|| = \sup_{\mathbf{X}} \frac{||\mathbf{A} \cdot \mathbf{X}|||}
{||\mathbf{X}||} .$$
$||\mathbf{A}||$ may be equivalently defined as the smallest upper bound of 
the eigenspectrum of $\mathbf{A}$.

The probability to have at least one resonance is thus directly related to 
the {\em norm} square $ ||\{\mu_q(0)\}||^2$ of the initial condition 
(this was also found numerically in \cite{SKKF09}). 
The probability $P_N \equiv 1-P_R$ to have 
{\em no resonance} thus has a {\em lower} bound, 
$P_N \geq 1-P_0 \kappa |\chi| \cdot  ||\mathbf{A}|| \cdot ||\{\mu_q(0)\}||^2$,
so that when the norm of the initial condition is small enough, 
$$ ||\{\mu_q(0)\}||^2 < \frac{1} {P_0 \kappa |\chi| \cdot  ||\mathbf{A}||} ,$$ 
we obtain $P_N>0$.
Thus, 
if we can prove that the norm $||\mathbf{A}|| $ is {\em not infinite}, 
the probability to have 
no resonance will be  {\em nonvanishing}, and  
go to $1$ when the norm of the initial condition goes to zero.

An upper bound for $||\mathbf{A}||$ can be obtained from 
\begin{eqnarray}
||\mathbf{A}|| \leq  \sup_q  \sum_{q^{\prime}} |A_{q,q^{\prime}}|  
\nonumber \\
 \leq  \sup_q \sum_{p\neq p^{\prime},q^{\prime}}  
\sum_{n} |\phi_n^{(p)}| \cdot | \phi_n^{(p^{\prime})}| \cdot |\phi_n^{(q)}|
\cdot |\phi_n^{(q^{\prime})}| 
\nonumber \\
< \sup_q \sum_{n}  (\sum_{p}|\phi_n^{(p)}|)^3 |\phi_n^{(q)}| .
\nonumber
\end{eqnarray}
If the eigenstates are exponentially localized,  
then $\sum_{p}|\phi_n^{(p)}| < +\infty$
and  $ \sum_{n}  |\phi_n^{(q)}| < +\infty$. If we assume these are bounded for 
all $n$ or
$q$ by the same constant $S$, then 
$||\mathbf{A}|| < S^4 < +\infty$.\footnote{If $\mu_q(0)$ is not arbitrarily 
chosen, we may get a better upper 
bound for the existence of KAM tori.
For example, if $\mu_q(0)= \delta_{q,q_0}||\{\mu_q(0)\}||$ is initially 
localized at a single Anderson mode $q^{\prime}=q_0$ \cite{FKS09},
we have 
$P_R  \leq  P_0 \kappa |\chi| (\sup_{q^{\prime}}  |A_{q^{\prime},q^{\prime}}|) 
\cdot ||\{\mu_q(0)\}||^2$
where $\sup_{q^{\prime}}  |A_{q^{\prime},q^{\prime}}| < ||\mathbf{A}||$. 
Thus, initial wave packets which are close to single Anderson modes 
survive much better as almost periodic solutions than those
which are arbitrarily spread (in Anderson space) at  the same norm. 
This effect is especially important when the localization length is large 
since then  $\sup_{q^{\prime}}  |A_{q^{\prime},q^{\prime}}| << ||\mathbf{A}||$.
The same is true if the initial wavepacket is split into several packets 
with smaller norm which are far apart 
at the scale of the localization length.}
Note that since $ \sum_{n}  |\phi_n^{(q)}|^2=1$,  if the localization length 
increases and diverges, then $ S > \sum_{n}  |\phi_n^{(q)}| \rightarrow \infty$. 
To obtain a reasonable estimate of $||\mathbf{A}||$ we 
assume an exponential bound for all
eigenstates, 
$$|\phi_n^{(p)}| < K \sqrt{\frac{1-\lambda^2}{1+\lambda^2}} \lambda^{|n-p|} ,$$
where $K$ is some constant and $ \lambda= e^{-1/\xi}$, 
where $\xi$ is the localization length.
Then 
$
S^4 =K^4 \frac{(1+\lambda)^6}{(1+\lambda^2)^2}\frac{1}{(1-\lambda)^2} .
$
%
When the localization length diverges at weak disorder we find 
$$||\mathbf{A}|| \lesssim  16 K^4  \xi^2 .$$ 
Consequently this upper bound for the norm  diverges, suggesting that
$||\mathbf{A}||$ might also diverge in the same way.

We did not consider the probabilities of resonances at higher orders 
$6,8,...$ which are
cumbersome to calculate. The correction on the bound of $P_R$ would be higher 
order in $ ||\{\mu_q(0)\}||^2$ and thus could be neglected in the limit of 
small norm. We conjecture 
that at each order $2p$, these probabilities can  also be bounded by 
convergent series multiplied with
$ ||\{\mu_q(0)\}||^{2(p-1)}$. 
The probability of higher order resonance is expected to decay exponentially 
with the order. Thus when the norm of the initial condition is not 
too large,
we would expect that the probability of  {\em no resonance at any order} 
is {\em non-vanishing} and  still bounded from below, 
going to 1 as  the norm goes to zero.

Note that for finite systems with size $N$,  the linear spectrum 
is always discrete and the series for 
bounding $P_R$ become finite sums, implying $||\mathbf{A}|| < +\infty$. 
Then, we know that the conclusion of our empirical argument is consistent 
with KAM theory, predicting the existence of $N-$dimensional invariant 
tori of quasiperiodic solutions at small enough amplitude 
(or equivalently, for finite systems, small enough norm)
with a probability going to $1$ at zero 
amplitude.
Our conjecture is that this argument also holds for infinite systems 
provided $||\mathbf{A}|| < +\infty$.
This situation occurs when the linear spectrum is purely {\em discrete} 
with exponentially localized eigenstates, 
but is not fulfilled when it contains an absolutely 
continuous part.
Then we would conclude, that  the norm region for initial conditions  
where KAM tori may exist 
shrinks to zero when the localization length diverges,
approaching the limit without disorder  
where the linear spectrum is absolutely continuous and 
no  $\mathit{l_2}$ almost periodic exact solution could survive 
due to radiaton.

Attempting to distinguish numerically KAM tori among other solutions 
in the random 
DNLS equation, we first consider that trajectories of KAM tori are almost 
periodic in time, and use the Harald Bohr theorem:

\textit{If a function $f(t)$ is almost periodic, then for any $\varepsilon>0$, 
there 
exists a relatively dense set of 
translations $\tau$ such that $|f(t)-f(t+\tau)|<\varepsilon$  for all 
$t \in ]-\infty, +\infty [$}. 
\textit{In other words, there exists a diverging 
subsequence $\tau_n$ fulfilling this 
property with $\tau_n<\tau_{n+1}$ 
and $\tau_{n+1}-\tau_n$ bounded}.

Thus, for KAM-like tori, {\em recurrences} should be observed numerically
in all quantities which depend on time 
such as local coordinates at arbitrary sites, momenta, 
participation number, etc.
Typically,  if recurrences 
appear for one of these quantities, they are found also for any other. 
Some problems with this method are obviously that 

(i) the Harald Bohr theorem can be checked only for finite times  
$\tau \in [0,T] $ and $t \in [0,T]$,  where  $T$  
is the time of integration,
 and for finite-size systems;

(ii) then $\varepsilon$ 
cannot be chosen too small for avoiding that recurrences 
become too rare,
and thus that  the corresponding pseudoperiod   exceeds  the integration time; 

(iii) the system size should be sufficient in order that the amplitude 
is practically zero at the edge, 
so that boundary effects can be neglected during the 
integration time $T$.

(iv) the integration accuracy should be good enough, for avoiding numerical
drift from KAM-like tori to neighboring chaotic trajectories. In practice, 
the relative error in the conserved quantities are kept at the order 
$10^{-6}$ or smaller in all simulations, with consistency checks for smaller 
systems reaching accuracies $10^{-8}- 10^{-10}$.  

If there are KAM tori persisting over infinite time, we should expect that 
their structure as some model parameter varies is
a {\em fat Cantor set with infinitely many gaps} 
due to resonances at all orders, but {\em nonvanishing Lebesgue measure} 
(or nonvanishing probability). 
Thus, one should find that the probability (in the space of initial 
conditions) that  recurrence persists
over a time $T$ does {\em not} shrink to zero as $T$ becomes very large.  
However, high order resonances (corresponding to small gaps) should manifest 
only after very long integration time. 
These trajectories may look almost periodic and exhibit recurrence over  
relatively long times,  before they blow up as chaotic trajectories.
(Such trajectories were numerically identified in \cite{SKKF09} as belonging 
to a ``regime I'' of rather small norm and/or strong disorder.)
Generally, all trajectories which remain recurrent after a time 
$T$ will be termed {\it T-recurrent}.

\begin{figure}
\begin{center}
\includegraphics[height=0.35\textwidth,angle=270]{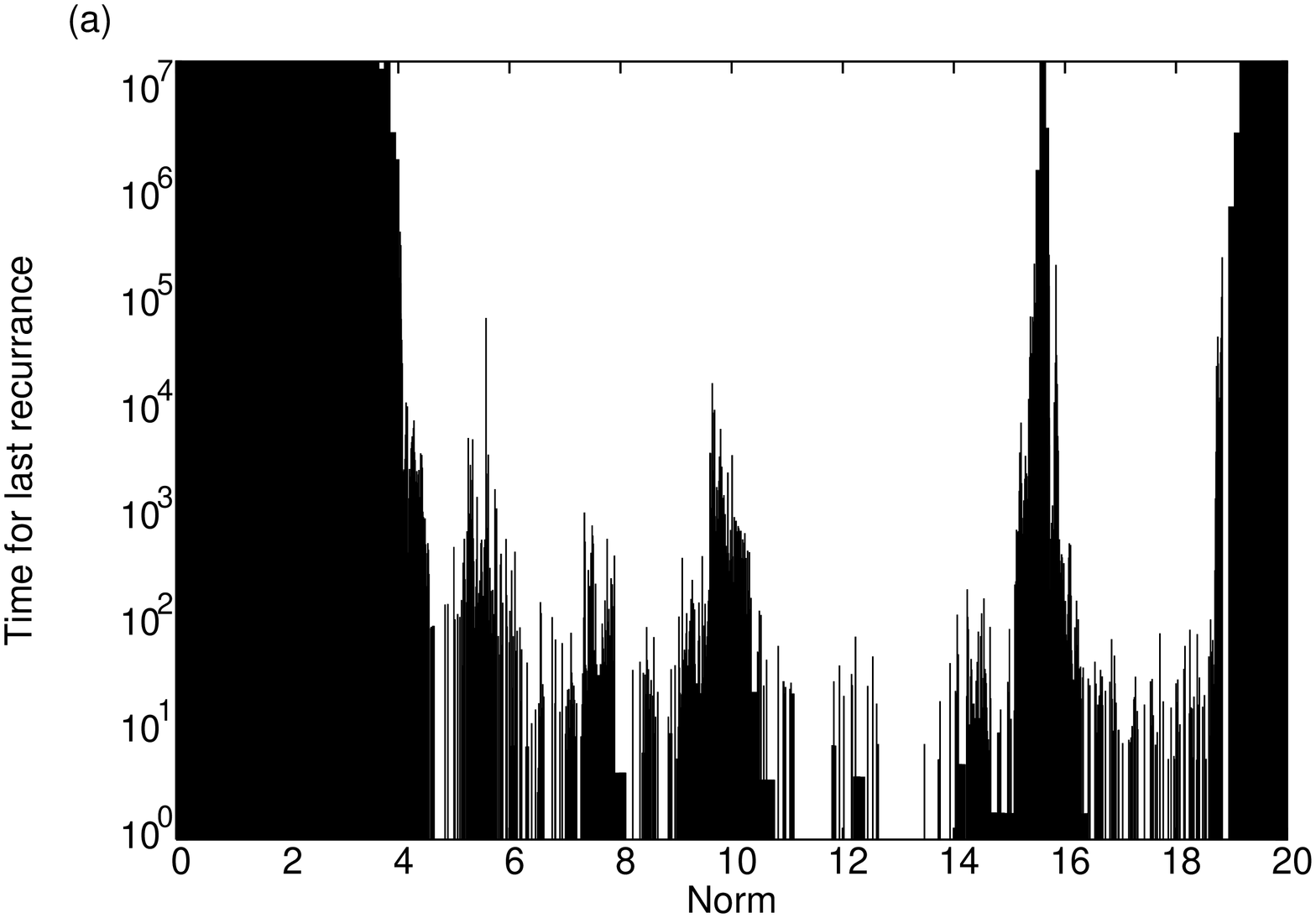}
\includegraphics[height=0.35\textwidth,angle=270]{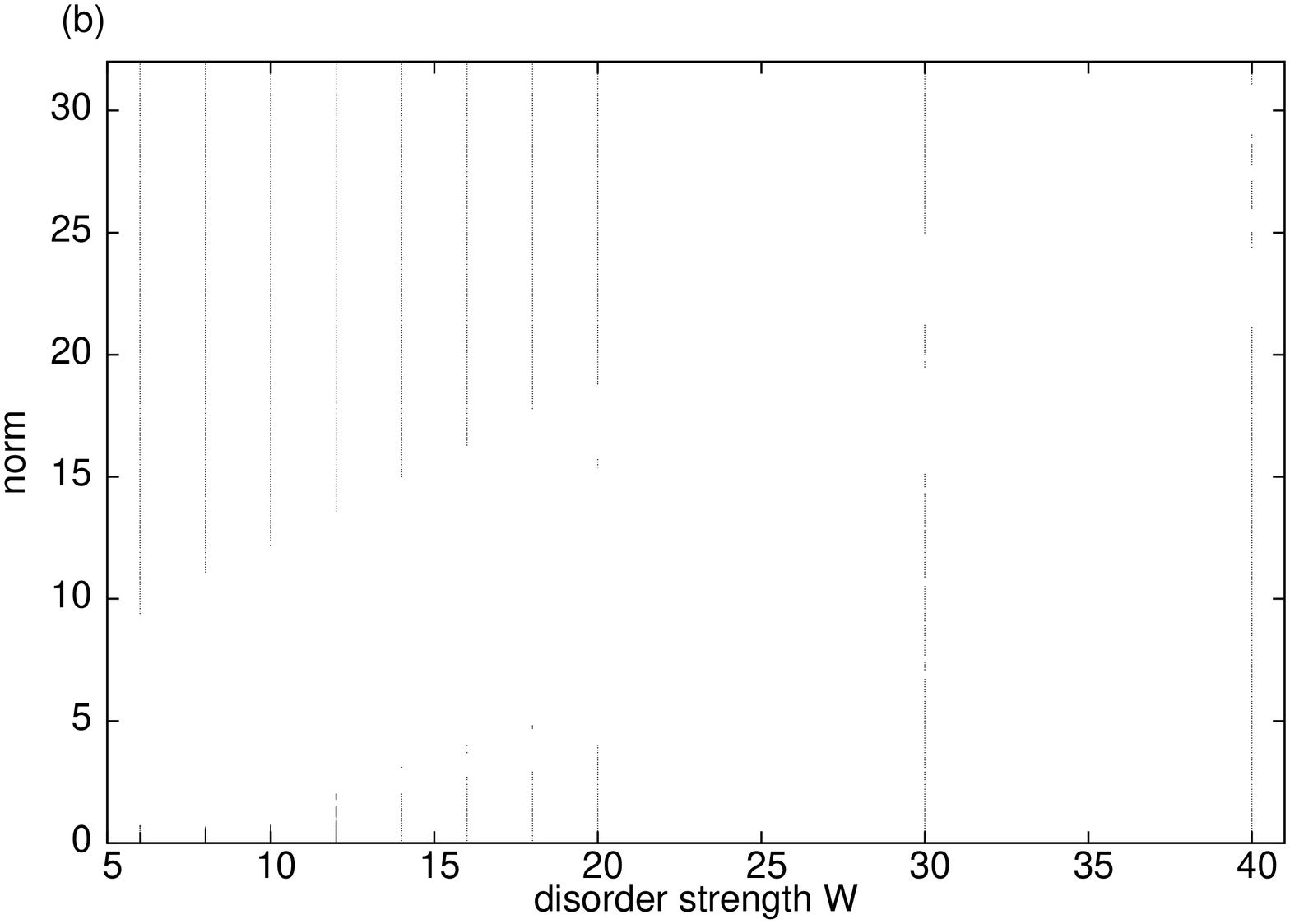}
\includegraphics[height=0.35\textwidth,angle=270]{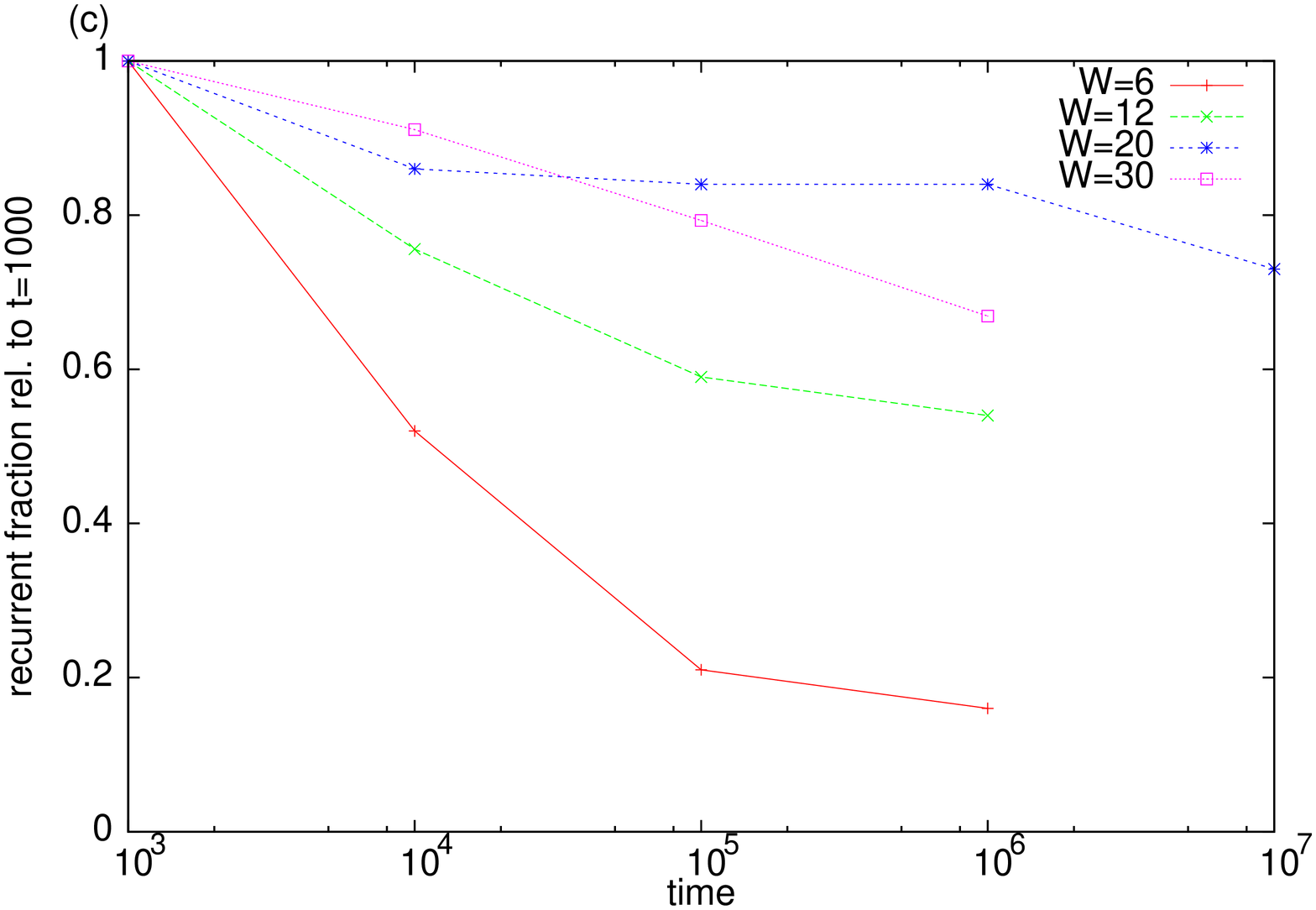}
\end{center}
\caption{(a) Last observed time for recurrence 
($\varepsilon / {\mathcal N} = 0.02$) in $|\psi_{n_0}|^2$ versus norm 
in a particular disorder realization with $W=20$. 
(b) Sets of $T$-recurrent trajectories at $T=10^4$ 
for various disorder strengths $W$. $\varepsilon / {\mathcal N}$ 
varied from 0.1 ($W=6$) to 0.005 ($W\geq 14$). 
(c) Fraction of the trajectories $T$-recurrent at $T=10^3$ 
(excluding the high-norm self-trapped regime) which remain 
  $T$-recurrent also at longer times, for various disorder strengths. 
At final integration times, $\varepsilon / {\mathcal N}$ is, respectively,
0.03 ($W=6$), 0.003 ($W=12$), 0.001 ($W=30$), and 0.0005 ($W=20$). 
In all figures, the disorder realization 
is the same, the initial condition 
$\psi_n(t=0) = \sqrt{{\mathcal N}}\delta_{n,n_0}$ with 
$\epsilon_{n_0}\approx 0.46529 W $, $C=1, \chi = -1$, and system size $N=500$.
 }
\label{fig_rec}
\end{figure}
An illustration is given in fig.~\ref{fig_rec}(a),  showing the 
last observed recurrence time as a function of norm for a single-site 
initial condition in a particular disorder realization for rather strong 
disorder $W=20$ (we chose 
the initial-site energy $\epsilon_{n_0}$ rather close to the upper band 
edge, in order that an increasing negative nonlinearity will ``scan'' most 
possible resonances inside the band). Any horizontal intersection of this 
graph at a given time $T$ yields the set of initial norm generating
$T-$recurrent trajectories. As can be seen, many trajectories 
remain
recurrent for times larger than $10^7$. In addition to the regime of small 
norm (here ${\mathcal N} \lesssim 3.6$) expected from our argument above, 
there is also a regime of recurrent 
states for ${\mathcal N} \gtrsim 19$, as well as a small interval around 
${\mathcal N} \approx 15.6$. The recurrent trajectories in the large-norm 
regime are related to the fact that, above some 
threshold norm, an increasingly larger part of 
the norm will self-trap around the initial site $n_0$ \cite{KKFA08}. 
The total norm available for the rest of the lattice will then actually 
decrease, and essentially the same argument as above could be used in 
support of a KAM-like regime. Note also that the simultaneous limit of 
strong disorder and large norm, 
$W \rightarrow \infty, {\mathcal N} \rightarrow \infty $ with 
$\chi {\mathcal N}/W$ finite, is equivalent 
(rescaling time)
to the anticontinuous (uncoupled) limit $C=0$  which is also integrable, and 
where the linear localization length is vanishing. In this limit, 
for any initial condition, each anharmonic oscillator $n$ exhibits a 
periodic motion with 
frequency $\omega_n^{\prime}\neq \epsilon_n$ different from the corresponding 
onsite linear frequency. We may thus expect a KAM-like regime 
also close to this 
limit.

Thus, though numerics cannot provide a rigorous proof, 
the most plausible interpretation is that
 that there is an underlying  Cantor set of initial conditions
generating KAM tori, persisting over infinite time and infinitely large 
systems. Note also that the gap structure of Fig. \ref{fig_rec}(a) is 
reminiscent of the ``stickiness'' phenomenon in low-dimensional systems, 
where many initial conditions close to 
(but outside) the Cantor set of KAM tori remain close for long times 
before they finally escape (compare, {\it e.g.}, with Figs.\ 7 and 10 in 
\cite{Cont97} for the standard map).

The variation of the sets of $T$-recurrent trajectories with the disorder
strength is illustrated by fig.~\ref{fig_rec}(b), for a rather modest time 
$T=10^4$. As predicted, there is always a small-norm regime where most 
trajectories are $T$-recurrent, the size of which grows with increasing 
disorder strength. There is also always a high-norm $T$-recurrent regime 
with lower boundary increasing with increasing disorder, 
since  the norm necessary for efficient self-trapping increases linearly 
with $W$ for a single-site initial 
condition \cite{KKFA08}. In-between these two regimes,  for larger 
$W$ there are also several intermediate regimes  of $T$-recurrence, separated 
by gaps with non-recurrent (chaotic) trajectories. For smaller $W$ the
relative sizes of these gaps grow, and they merge into one single main gap
of trajectories which typically show a chaotic time-evolution and 
spread sub\-diffusively
\cite{Shep93,Mol98,PS07,SKKF09} (for long but possibly finite times).  

Similar pictures as fig.~\ref{fig_rec}(b) can be obtained for larger times, 
although obtaining a good resolution with sufficient numerical accuracy to 
clearly identify recurrences makes it very time-consuming for times larger
than $\sim 10^6$. As a rule of thumb, to determine persistent recurrences
$\varepsilon$ is divided by two for each 
order of magnitude in time.
In fig.~\ref{fig_rec}(c), we give an example showing how the 
measure of the set of $T$-recurrent initial conditions 
varies with $T$, for  different
strengths $W$ of the same 
disorder realization. 
In order to obtain a finite set, we here exclude recurrent 
trajectories belonging to the high-norm (self-trapped) regime, and moreover 
for comparison we normalize the sets by dividing with the number of 
$T$-recurrent trajectories at $T=10^3$.   
The data of  fig.~\ref{fig_rec}(c)  suggest the existence of an asymptotic 
set with a nonvanishing measure at infinite time. However,
for stronger disorder, it is clear from fig.~\ref{fig_rec}(c) that to 
get a clear picture of the asymptotic measure of this set, considerably longer 
integration times would be needed. This is due to the fact that the
fraction of ``sticky'' trajectories, which are $T$-recurrent over very large
but finite times, increases with the disorder strength 
(cf.\ fig.~\ref{fig_rec}(a)). 

Thus, the numerical study of $T$-recurrent trajectories for finite times 
can give us only an indication about the true nature of the KAM-like 
trajectories, and in particular the presence of long-time 
``sticky'' trajectories 
makes it extremely difficult, {\it e.g.}, 
to resolve the Cantor-set structure of 
resonances in the low-norm regime within a reasonable amount of computer 
time. We therefore now turn to discusss another technique to numerically 
distinguish KAM tori, in terms of the {\em tangent map} and the corresponding 
finite-time Lyapunov exponents. 
Small perturbations of eq.~(\ref{rdnls}) yield the Hill equation
\begin{equation}
 i \dot{\eta}_n = ( \epsilon_n  + 2 \chi  |\psi_n|^2) 
\eta_n+\chi  \psi_n^2 \eta_n^{\star} - C(\eta_{n+1}+\eta_{n-1}) .
 \label{rdnlstg}
 \end{equation}
If eq.~(\ref{rdnls}) possesses an almost periodic solution with the discrete 
set of frequencies 
$\omega_1,\omega_2,...,\omega_p,...$,
corresponding to  KAM tori with full 
dimension, 
$$\psi_n =F_n(\omega_1 t+\alpha_1,\omega_2 t+\alpha_2,...,
\omega_p t+\alpha_p,...; \omega_1,\omega_2,...,\omega_p,...) ,$$ 
where $F_n(x_1,x_2,...x_p,...;\omega_1,\omega_2,...,\omega_p,...)$ is 
$2\pi$-periodic with respect to 
$x_1, x_2, ..., x_p,...$, then $\eta_n(t) = \frac{\partial F_n}{\partial x_p}$ is an 
almost periodic solution of eq.~(\ref{rdnlstg}).
One gets a complete basis of solutions of eq.~(\ref{rdnlstg}) by adding also 
the solutions
$\eta_n(t) = \frac{d F_n}{d \omega_p}=
t  \frac{\partial F_n}{\partial x_p} + \frac{\partial F_n}{\partial \omega_p}$, 
where also  $ \frac{\partial F_n}{\partial \omega_p}$ is almost periodic.
Consequently, if the solution of eq.~(\ref{rdnls}) corresponds to  a 
KAM torus, the general solution of eq.~(\ref{rdnlstg})
grows {\em linearly} as a function of 
time. Note also that for any solution 
$\psi_n$ to eq.~(\ref{rdnls}), there is a
trivial solution 
$\eta_n=i \psi_n$ to eq.~(\ref{rdnlstg}) corresponding to a
global phase rotation. Numerically, we remove this component 
by subtracting the projection of $\eta_n(t)$, obtained by 
integrating a
randomized initial condition $\eta_n(0)$, on this vector.

\begin{figure}
\begin{center}
\includegraphics[height=0.39\textwidth,angle=270]{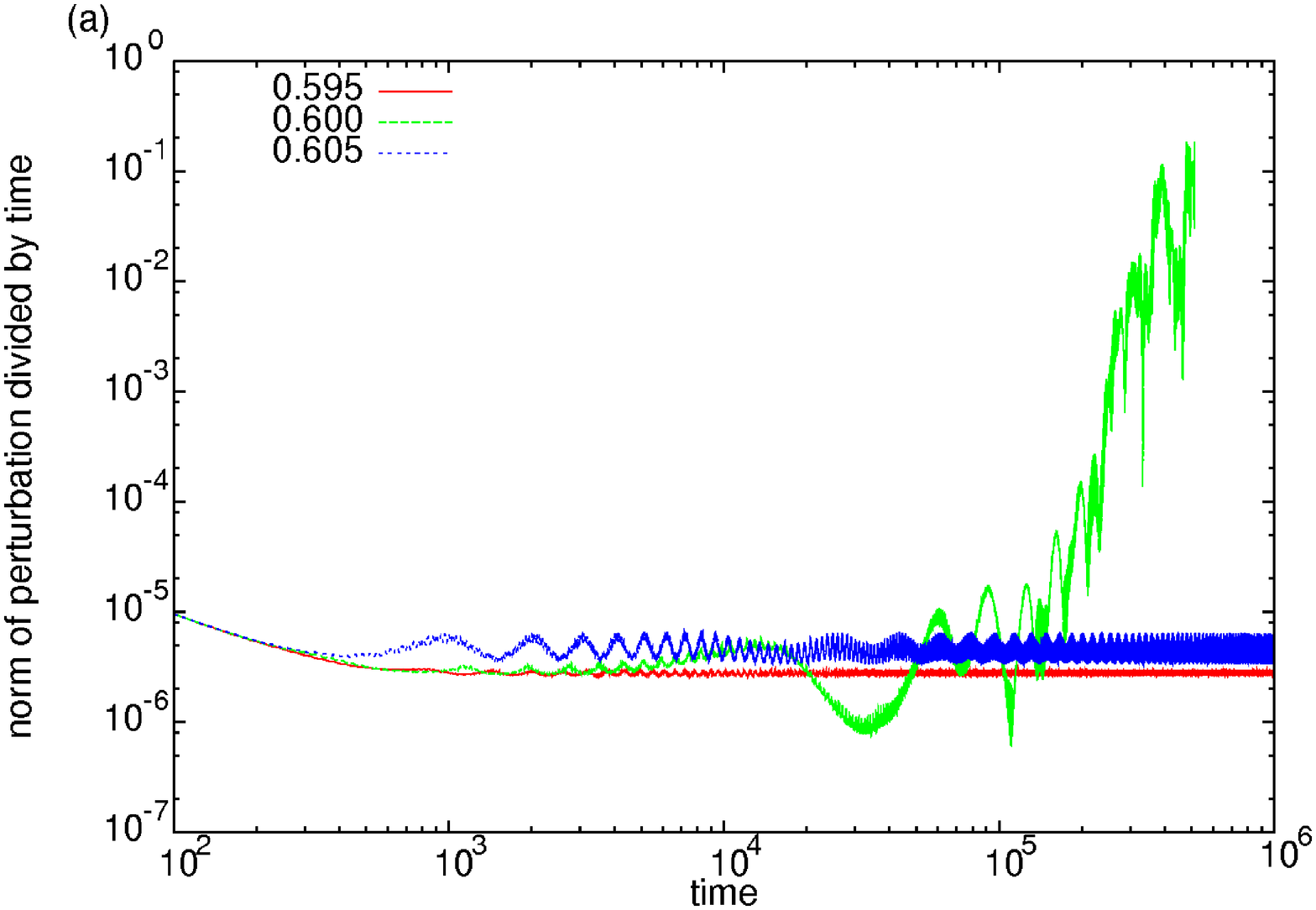}
\includegraphics[height=0.39\textwidth,angle=270]{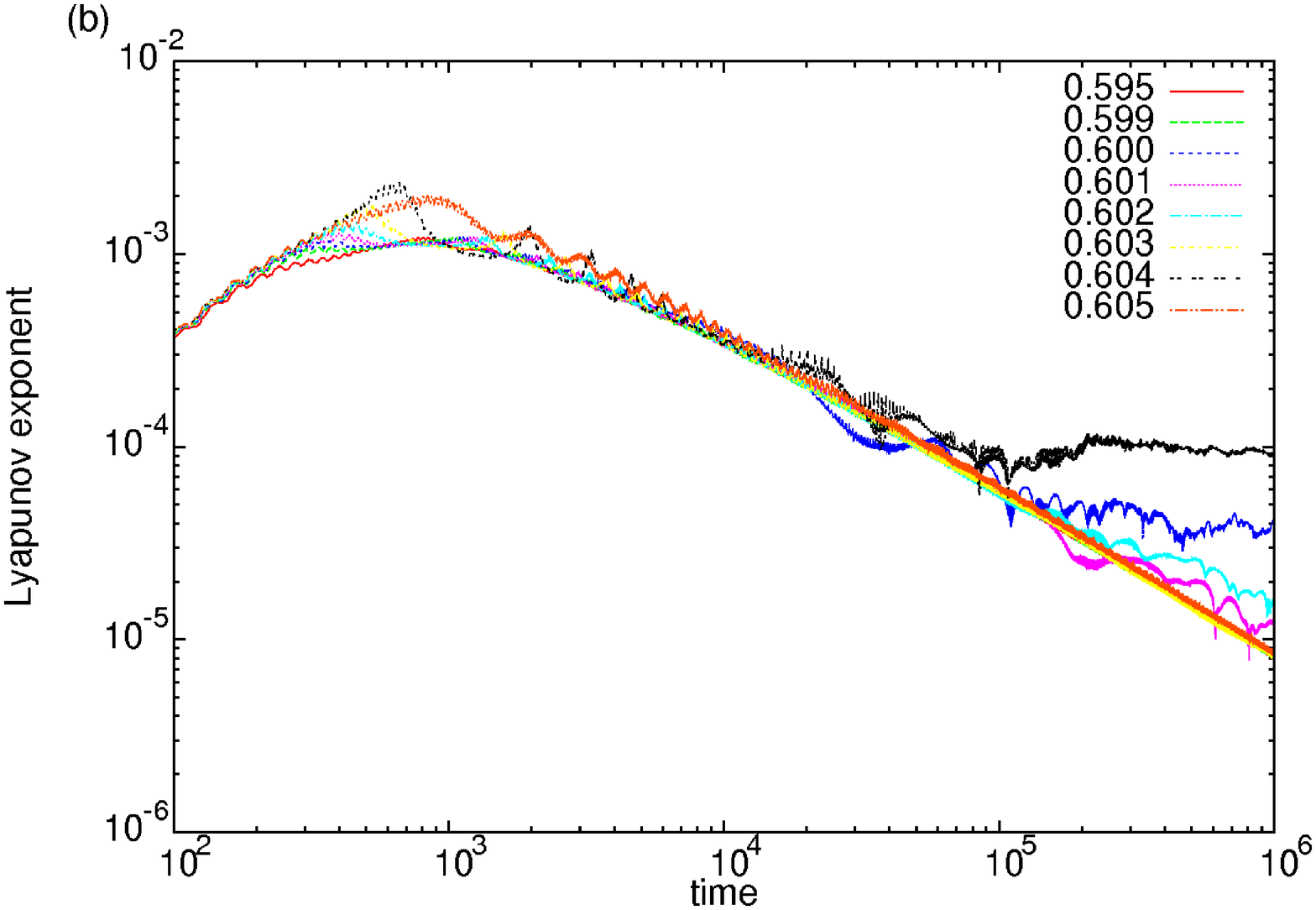}
\end{center}
  \caption{(a) Total norm of perturbation $\eta_n(t)$ divided 
by time, for three solutions corresponding to single-site initial conditions 
$\psi_n$
with slightly different ${\mathcal N} \approx 0.6$. 
(b) Finite-time Lyapunov exponents for 8 solutions in the same 
regime as in (a).  At time 
$10^6$, the upper curves in (b) correspond, from top to bottom, to 
${\mathcal N}=0.604, 0.600, 0.602, 0.601$, while the lower curves 
for ${\mathcal N} = 0.595, 0.599, 0.603, 0.605$ all follow very closely
a curve $\sim \log t / t$, as expected for KAM tori. 
Disorder strength 
$W=12$, other parameters and disorder realization same as in 
fig.~\ref{fig_rec}.
}
\label{fig_tdiv}
\end{figure}

Thus, considering the total norm of the perturbation divided by $t$, 
$\frac{1}{t}||\eta(t)||$ where $||\eta||= \sqrt{\sum_n|\eta_n|^2}$, 
this quantity will exhibit bounded 
oscillations for all times if $\psi_n$ corrsponds to a KAM torus, 
and diverge exponentially 
with a positive Lyapunov exponent for  
any chaotic trajectory. A numerical illustration is given in 
fig.~\ref{fig_tdiv}(a), 
showing a narrow resonance gap in the low-amplitude 
KAM-like regime around ${\mathcal N} = 0.600$ 
for disorder strength $W=12$. 

From the so obtained  numerical solutions  $\eta_n(t)$, we calculate 
finite-time Lyapunov exponents as 
$\Lambda(t_M)=\frac{1}{t_M}\sum_{m=1}^M\log{\frac{ || \eta(t_m)||}
{ || \eta(t_{m-1})||}}$, where $t_0=0$, and $t_m$ are chosen to correspond to 
a Poincar\'e section defined by $|\psi_{n_0} (t)|^2$ 
having a local maximum at each $t=t_m$. Thus, for a recurrent trajectory, 
the optimal recurrence times $\tau_k$ form a subset of the $t_m$. Moreover, 
for an almost-periodic trajectory with a time-linear growth of  
$||\eta||$, $\Lambda(t)$ should decrease asymptotically to zero as  
$\Lambda(t) \sim \log(t)/t$ for large times. 
As can be
seen from the example in fig.~\ref{fig_tdiv}(b), 
the behaviour of $\Lambda(t)$ 
in the neighborhood of sharp resonances is very sensitive to small 
parameter variations. In this example, the trajectories are seen to 
be (possibly weakly) chaotic in the 
interval $0.600 \leq {\mathcal N} \leq 0.602$ and in an even narrower 
interval around ${\mathcal N} \approx 0.604$, while apparently almost-periodic
trajectories (with no visible deviations from the asymptotic behaviour 
$\Lambda(t) \sim \log(t)/t$ for times larger than $10^6$)  are seen e.g.\ 
for ${\mathcal N} = 0.595, 0.599, 0.603$ and 0.605. Thus, this again supports 
the existence of a finite-measure Cantor set of almost-periodic KAM tori.

Comparing the numerical results from the analysis of recurrences and tangent 
map, they are consistent in the sense that when recurrences are 
lost, there is a clear deviation in $||\eta||$ from time-linear growth, and 
in most cases the growth is exponential with well-defined non-zero $\Lambda(t)$ 
(at least for long times). However, generally (and in particular 
for strong disorder with many ``sticky'' trajectories), the 
tangent-map criterion is considerably more sensitive, and may signal 
a chaotic behaviour several orders of magnitude in time before recurrence 
is finally  lost.  

\begin{figure}
\centerline{\includegraphics[height=0.30\textwidth,angle=270]
{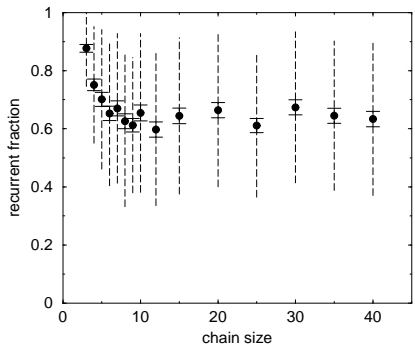}}
  \caption{Ensemble-averaged fractions of trajectories $T$-recurrent
at $T=2\cdot 10^5$ versus chain size, for random single-site initial conditions 
with $0 < {\mathcal N} \leq 20$ and disorder strength $W=12$. 100 different
realizations were used for each size, and dashed lines show the
standard deviations of the distributions. 
Trajectories were considered $T$-recurrent if $\Lambda(T) < 10^{-4}$ 
(cf.\ fig.~\ref{fig_tdiv}).
}
\label{fig_finite}
\end{figure}

Generally, we find the KAM-like trajectories to be exponentially 
localized, decaying essentially with the largest 
linear localization length 
$\xi$ ({\it e.g}, $ \xi_\mathrm{max} \approx 1.1$  for $W=12$ as in 
fig.~\ref{fig_tdiv}). Thus we may expect that, for stronger disorder, 
the KAM-like trajectories observed for large systems should be only 
slight perturbations of true KAM tori existing for short chains. This was also 
confirmed numerically. An illustration is given in 
fig.~\ref{fig_finite} where, for $W=12$ and single-site initial 
conditions 
$\psi_n(0) = \sqrt{{\mathcal N}}\delta_{n,n_0}$ 
with $0 < {\mathcal N} \leq 20$, we calculate the fraction of the 
total number of trajectories remaining $T$-recurrent after 
$T=2\cdot 10^5$, for chain lengths with $3 \leq N \leq 40$. For each chain 
length, we used 100 different (independent) disorder realizations, and 
determined ensemble averages and standard deviations (note that for 
small chains,
ensemble statistics can be done with reasonable computational effort). 
As can be seen, 
already for $N=6$ ($ \approx 5 \xi_\mathrm{max}$) this fraction has essentially 
converged to its large-chain limit. The large spread between different 
realizations is essentially due to the fact that $\epsilon_{n_0}$ also 
is chosen randomly: if $\epsilon_{n_0}$ is close to the linear band bottom,  
self-trapping occurs already for small values of $ {\mathcal N}$, 
and therefore most trajectories will be KAM-like. By contrast, 
realizations with
$\epsilon_{n_0}$ close to the band top as in 
figs.~\ref{fig_rec}-\ref{fig_tdiv} should yield the largest amount of 
resonances. Indeed, for the particular realization used for the 
large-size simulations  in 
figs.~\ref{fig_rec}-\ref{fig_tdiv}, we obtain for $W=12$ 
a total fraction of $38\%$ 
$T$-recurrent trajectories in the interval   $0 < {\mathcal N} \leq 20$ after 
$T=10^5$ (decreasing to  $37\%$ at $T=10^6$), essentially coinciding with 
the lower bounds in  fig.~\ref{fig_finite} ({\it i.e.}, this realization 
does not behave exceptionally). 

In summary, our empirical and numerical arguments suggest that in the 
infinite random DNLS model (\ref{rdnls}), 
there are two kinds of initial wavepackets 
both occuring with nonzero probability: (i) those generating 
spatially localized (non-spreading),
almost 
periodic solutions (KAM tori); 
and (ii) wave-packets which are initially chaotic and 
spreading \cite{Shep93,Mol98,PS07,SKKF09}. Our results should stimulate 
further attempts towards more rigorous treatments, as well as more detailed 
numerical studies of the role of such tori in various situations. 
Ongoing work suggests, {\it e.g.}, that also large-norm wavepackets extended 
on many sites could generate KAM tori with high probability if the norm 
density is small enough. Thus, the spreading of chaotic wavepackets may after 
long times drastically slow down, if they become sticky to such tori. 
The KAM-like trajectories may also be experimentally observable, 
{\it e.g.}, in disordered waveguide arrays \cite{Lahini}, or in disordered 
bosonic systems \cite{BEC}.

\begin{acknowledgments}

We thank S.\ Flach, D.\ Krimer, Ch.\ Skokos and 
J.\ Bodyfelt for useful suggestions. 
MJ acknowledges 
support from the Swedish Research Council. 

\end{acknowledgments}

\end{document}